\documentclass[nofootinbib]{revtex4}
\usepackage{graphicx}
\usepackage{amsmath}
\def\be{\begin{equation}}
\def\ee{\end{equation}}
\def\ve{\varepsilon}
\def\GeV{{\rm\ GeV}}

\begin{document}

\title{Proton charge and magnetic rms radii
from the elastic $ep$ scattering data}

\author{Dmitry~Borisyuk%
}

\affiliation{Bogolyubov Institute for Theoretical Physics,
Metrologicheskaya street 14-B, 03680, Kiev, Ukraine}

\begin{abstract}
The elastic electron-proton scattering data are analysed in order to
determine proton charge and magnetic rms radii, $r_E$ and $r_M$.
Along with the usual statistical error, we try to estimate
a systematic error in the radii, caused by the inadequacy of
particular form factor parameterization employed.
The range of data to use in the analysis is chosen so as to
minimize the total (statistical + systematic) error.
We obtain $r_E = 0.912 \pm 0.009 {\rm\ (stat)} \pm 0.007 {\rm\ (syst)\ fm}$
and $r_M = 0.876 \pm 0.010 {\rm\ (stat)} \pm 0.016 {\rm\ (syst)\ fm}$.
The cross-section data were corrected for two-photon exchange.
We found that without such corrections
obtained $r_E$ and $r_M$ are somewhat smaller
while the quality of fit is worse.
\end{abstract}

\maketitle

\section{Introduction}

 Charge and magnetic rms radii are important low-energy
 characteristics of proton, showing its ``size''.
 They are defined by
\be \label{radii}
 r_E^2 = - 6 \left. \frac{dG_E}{dQ^2}\right|_{Q^2=0}
  \qquad
 r_M^2 = -\frac{6}{\mu}  \left. \frac{dG_M}{dQ^2}\right|_{Q^2=0}
\ee
 where $G_E$ and $G_M$ are electric and magnetic form factors (FFs)
 and $\mu = G_M(0)$ is proton magnetic moment.
 In the nonrelativistic approximation the radii can be expressed via moments
 of charge and magnetization densities:
\be
 r_E^2 = \smallint \rho_E(r) r^2 dV, \qquad r_M^2 = \smallint \rho_M(r) r^2 dV
\ee
 The knowledge of proton radii is necessary for the calculation of
 other low-energy observables with taking finite proton size into account.
 The examples are Lamb shift and hyperfine splitting
 in hydrogen, where the experimental errors are several orders
 less than the proton finite size corrections \cite{LowE}.
 
 Proton radii are usually determined from the elastic
 electron-proton scattering data. One chooses
 some FF parameterization and then performs a fit
 to experimentally measured FFs or cross-sections.
 The radii are then obtained by Eq.(\ref{radii}).
 Nevertheless, this procedure is somewhat ambiguous.
 At first, there are many ways to parameterize FFs.
 The most fundamental way is to construct a parameterization
 which has proper analytic structure (poles and cuts in the time-like region,
 asymptotic behaviour, etc.) \cite{MMD,MMD2}.
 The drawbacks of this method are very complicated expressions for FFs,
 need for a model input, and need to analyze proton and neutron FFs
 simultaneously and for all $Q^2$.
 If we are interesting just in proton radii,
 it is simpler to restrict oneself to electron-proton scattering
 at small $Q^2$, as we do in the present work.
 The FF parameterization is not constrained anyhow;
 it may be chosen for the reasons of convenience and accuracy of results.
 The examples are linear, quadratic \cite{quadParam},
 continued-fraction \cite{Sick} parameterization, etc.
 
Our work is most similar to Ref.~\cite{Sick}.
As it was done in Ref.~\cite{Sick}, we parameterize electric and magnetic FFs
simultaneously and fit the cross-section data directly, instead of performing
Rosenbluth separations and fitting obtained FFs.
However, we use another, more intuitively understandable, FF parameterization,
 normalize the data from each experiment independently
 (the normalization factors become fit parameters),
 and evaluate both electric and magnetic radii. 

Besides the cross-section data, today there is a number of
data on polarization observables. However we do not use the latter
in our analysis, because the polarization observables
do not depend on $G_E$ and $G_M$ separately,
but depend on the $G_E/G_M$ ratio only.
The data on $G_E/G_M$ can constrain a small difference $r_E^2 - r_M^2$,
but not $r_E$ or $r_M$ itself. Additionally, there are about
10 polarization data points for $Q^2 < 1 \GeV^2$,
to be compared with hundreds of cross-section points available.
Therefore the effect of polarization data inclusion would be extremely weak.

The distinctive feature of our work
is the new approach to the error estimation.
Usually, two sources of the error are considered:
1) the statistical errors in the individual data points,
and
2) the overall normalization uncertainty for each experiment.
But there is third error source:
the inadequacy of chosen FF parameterization.
Indeed, if we try to use, for instance, linear parameterization for
$0 < Q^2 < 1\GeV^2$, we will never obtain accurate results,
since the FFs are very different from a linear function
on this interval.
On the other hand, choosing too small $Q^2$ range increases
first and second type errors,
since less data points will be used and FF value will be too close to 1.
The latter circumstance was also pointed out in Ref.~\cite{Sick},
but to our opinion, author did not treat the third type of error properly.
In particular, author studies the scatter of $r_E$,
associated with the choice of fit type
(i.e. number of terms in the parameterization and range of data
to use), and counts it as a part of the statistical error.
 Instead, we assert
   that fit type should be chosen so as to minimize the total error,
   which consists of three above-mentioned parts.
 Therefore the errors in Ref.~\cite{Sick} turn out to be overestimated.
The detailed discussion of the error estimation is given in Sec.~\ref{Sec:Err}.

 The importance of two-photon exchange (TPE) corrections
 for the proton radius extraction
 was several times stressed in the literature.
 We apply TPE corrections to the cross-sections
 before fitting and check their influence on the results.
 
\section{Fitting procedure}\label{Sec:Proc}
 The following $\chi^2$ function is minimized:
\be \label{chi2}
 \chi^2 = \sum_{e,i}
  \left(
   \frac{\sigma^{\rm exp}_{e,i}-N_e \sigma^{\rm th}_{e,i}}{d\sigma_{e,i}}
  \right)^2
 + \sum_e \left( \frac{N_e-1}{dN_e} \right)^2, 
\ee
where $e$ enumerates experiments, $i$ --- data points within each experiment,
$\sigma^{\rm exp}_{e,i}$ and $d\sigma_{e,i}$ are experimental cross-sections
and their point-to-point errors, $N_e$ are normalization factors
(to be determined by fitting), $dN_e$ are quoted normalization uncertainties
and $\sigma^{\rm th}$ is theoretical cross-section:
\be \label{sigma}
 \sigma^{\rm th} = \ve G_E^2(Q^2) + Q^2 G_M^2(Q^2)/4M^2 + \sigma^{2\gamma}(Q^2,\ve).
\ee
Here $\sigma^{2\gamma}(Q^2,\ve)$ is TPE correction,
calculated according to Ref.~\cite{ourBox}.
Since we are interesting in proton radii,
which are related to FF behaviour near $Q^2=0$,
it is enough to take into account only the elastic intermediate
state in the calculations of TPE. The inclusion of TPE corrections
is important, since they not only do not vanish but grow rapidly as $Q^2\to 0$,
what may noticeably affect the radius extraction \cite{ourLow}.

The FFs were parameterized as
\be
  G_E(Q^2) = (1-\xi/\xi_0)^2 \sum_{k=0}^n a_k \xi^k, \qquad
  G_M(Q^2) = (1-\xi/\xi_0)^2 \sum_{k=0}^{n-1} b_k \xi^k,
\ee
where $\xi = Q^2/(1+Q^2/\xi_0)$ with $\xi_0 = 0.71 \GeV^2$.

The multiplier in front of the sum is equal to $(1+Q^2/\xi_0)^{-2}$ and
represents dipole falling of FFs,
thus the remaining polynomial should be approximately constant
over the full $Q^2$ range; this reduces statistical errors in its coefficients.
The polynomial in $\xi$ is better than the polynomial in $Q^2$,
since $\xi$ remains finite at $Q^2\to\infty$,
thus the behaviour of polynomial at higher $Q^2$ is well-controlled,
whereas the polynomial in $Q^2$ inevitably grows with $Q^2$.
We use the polynomial of degree $n$ for parameterization of $G_E$ and
$n-1$ for $G_M$, because the contribution of $G_M$
to the cross-section is damped by $Q^2/4M^2$
[see Eq.~(\ref{sigma})], and thus the $G_M$ form factor
is less significant at low $Q^2$.
The choice of the parameter $n$ is discussed in Sec.~\ref{Sec:FType}.

The proper FF normalization is ensured by keeping
$a_0 = G_E(0) = 1$ and $b_0 = G_M(0) = 2.793$ fixed.
So the free fit parameters are coefficients
$a_1$, ..., $a_n$; $b_1$, ..., $b_{n-1}$ and normalization factors $N_e$.
After the fitting, $r_E$ and $r_M$ are calculated via Eq.(\ref{radii}).

\section{Error estimation}\label{Sec:Err}
The statistical error can be computed
in the straightforward way (see exact formulae below),
and includes both the error caused by point-to-point uncertainties
and the error caused by an uncertainty in the normalization of each experiment.
Some authors \cite{Sick} identify the last one as a systematic error,
but we do not agree with them.
While the normalization errors are systematic with respect
to a single experiment data, they become statistical with respect
to the proton radius, which is extracted from a number of experiments,
every one having its own normalization. Moreover, if an experiment is re-done
from scratch, the normalization coefficient will be different each time,
similarly to statistical errors.

Instead, we recognize another source of really systematic error in the radius.
Suppose for simplicity that FF is measured directly and we fit it with,
say, quadratic polynomial.
If we knew {\it a priori } (by some theoretical arguments)
that FF in any case must have this definite form%
\footnote{ For example, if instead of FFs
  we were studying time dependence of radioactive decay rate,
  we knew that it must have exponential form.
  So in this case we should fit the data with exponential,
  not polynomial or the like.}%
, the only thing to do
was to extract the coefficients of the polynomial, and the error in the FF
would be the consequence of the errors in these coefficients.
But actually we do not know anything about FFs {\it a priori }
(except the normalization at $Q^2=0$).
Therefore even if (accidentally) the experimentally measured FF values
coincide with the true ones,
the FF obtained by the fitting procedure will deviate
from the true one, simply because the assumed parameterization
cannot reproduce true FF well enough.
The resulting difference has all properties of
a systematic error: it will not change if the experiments are re-done
and it does not depend on the precision of the data.

From the mathematical point of view,
the radius $r_{\rm fit}$, yielded by the fitting procedure,
is a function of experimentally measured cross-sections $\sigma_{e,i}^{\rm exp}$,
\be
 r_{\rm fit} = r_{\rm fit}(\sigma_{e,i}^{\rm exp})
\ee
The measured cross-sections deviate from the true ones, thus
\be
 \sigma_{e,i}^{\rm exp} =
   (1+\Delta N_e) \sigma_{e,i}^{\rm true} + \Delta \sigma_{e,i} 
\ee
where $\Delta N_e$ and $\Delta\sigma_{e,i}$ are unknown random quantities.
The squares of quoted statistical and normalization errors,
$d\sigma_{e,i}$ and $dN_e$, used in Eq.(\ref{chi2}),
are variances of these quantities:
\be
 d\sigma_{e,i}^2 = \overline{\Delta\sigma_{e,i}^2},\ \ \ 
 dN_e^2 = \overline{\Delta N_e^2}
\ee
(the bar denotes mathematical expectation).
Since $\Delta \sigma$ and $\Delta N$ are small enough,
we may expand $r_{\rm fit}$ into Taylor series about $\sigma_{e,i}^{\rm true}$:
\be \label{rfit}
 r_{\rm fit} = r_0 + \sum_{e,i} c_{e,i}
     [ \Delta N_e \sigma_{e,i}^{\rm true} + \Delta\sigma_{e,i} ]
  + \sum_{e,i,e',i'} d_{e,i;e',i'}
     [ \Delta N_e \sigma_{e,i}^{\rm true} + \Delta\sigma_{e,i} ]
     [ \Delta N_{e'} \sigma_{e',i'}^{\rm true} + \Delta\sigma_{e',i'} ]  
  + O(\Delta\sigma,\Delta N)^3,
\ee
where
\be
r_0      = r_{\rm fit}(\sigma^{\rm true}), \qquad
c_{e,i}  = \frac{\partial r_{\rm fit}}{\partial\sigma_{e,i}^{\rm exp}} (\sigma^{\rm true}), \qquad
d_{e,i;e',i'} = \frac{1}{2}
  \frac{\partial^2 r_{\rm fit} }
       { \partial\sigma_{e,i}^{\rm exp}\partial\sigma_{e',i'}^{\rm exp} }
  (\sigma^{\rm true}).
\ee
The quantity $r_0$ in general {\it is not equal} to true radius $r_{\rm true}$;
the difference is what we call systematic error:
\be
 dr_{\rm syst} = r_0 - r_{\rm true} = \overline{r_{\rm fit} - r_{\rm true}}
\ee
The total error of the radius is naturally defined by
\be
  dr^2 = \overline{(r_{\rm fit} - r_{\rm true})^2}
\ee
Using Eq.(\ref{rfit}) and assuming, as usual,
no correlation between different data points and experiments,
\be
\overline{\Delta\sigma_{e,i}\Delta\sigma_{e',i'}}
  = \delta_{ee'}\delta_{ii'}d\sigma_{e,i}^2, \qquad
\overline{\Delta N_{e} \Delta N_{e'}} = \delta_{ee'} dN_e^2,
\ee
we easily obtain
\be
 dr^2 = (r_0-r_{\rm true})^2 + 
 \sum_{e,i,i'} [(r_0 - r_{\rm true})d_{e,i;e,i'} + c_{e,i} c_{e,i'}] 
 [\sigma_{e,i}^{\rm true} \sigma_{e,i'}^{\rm true} dN_e^2
   + \delta_{ii'} d\sigma_{e,i}^2] + O(dN,d\sigma)^4.
\ee
If $dr_{\rm syst}$ is reasonably small, then the first term in the square bracket
can be neglected;
also $\sigma^{\rm true}$ can be safely replaced by $\sigma^{\rm exp}$,
thus we obtain
\be
 dr^2 = dr_{\rm syst}^2 + dr_{\rm stat}^2,
\ee
where 
\be
 dr_{\rm stat}^2 = \sum_{e,i,i'} c_{e,i} c_{e,i'}
  [\sigma_{e,i}^{\rm exp} \sigma_{e,i'}^{\rm exp} dN_e^2
    + \delta_{ii'} d\sigma_{e,i}^2]
\ee
is the statistical error.

The systematic error is not so easy to estimate as the statistic one.
It follows from Eq.(\ref{rfit}), that the systematic error is
the error of the radius that would be obtained if
$\sigma^{\rm exp}$ coincide with $\sigma^{\rm true}$.
Thus to estimate it we need to know true FFs, but this is exactly
what is unknown. The best thing we can do is to use pseudo-data.
First, we choose some realistic FF parameterization
(from one of Refs.~\cite{Sick,MMD,paramK,paramB,paramA} or dipole fit).
After that $r_{\rm true}$ is known. Then we generate pseudo-data
at $\ve$ and $Q^2$ of experimental data, assign them the same uncertainties
$d\sigma_{e,i}$, but no errors ($\Delta\sigma_{e,i}=0$, $\Delta N_e=0$).
Then the fitting procedure, described in Sec.~\ref{Sec:Proc},
is applied and the radius $r_{\rm fit}$ is obtained.
The systematic error is the difference $r_{\rm fit}-r_{\rm true}$,
since the statistical errors are absent by construction (of course its value
depends on the parameterization chosen). The maximal absolute value among
different parameterizations is taken as an estimate of the systematic error.
\section{Choice of fit type}\label{Sec:FType}
Before we perform the fitting of actual data, we must choose the polynomial
degree $n$ and the higher bound for momentum transfer $Q_{\rm max}^2$
(a data point will be included in the analysis if it has $Q^2\le Q_{\rm max}^2$).
We will show that both $n$ and $Q_{\rm max}^2$ can be chosen to minimize
the total uncertainty of the radius.

Suppose that the polynomial degree $n$ is fixed and $Q_{\rm max}^2$ is varied.
The statistical error will decrease as $Q_{\rm max}^2$ increases,
simply because more data points are used. 
On contrary, the systematic error will grow: the polynomial of some fixed
degree may be a reasonable approximation of FF on a small interval of $Q^2$,
but becomes less adequate on a larger interval, because the actual FF
is not a polynomial at all.

\begin{figure}[h]
\centering
 \includegraphics[width=0.45\textwidth]{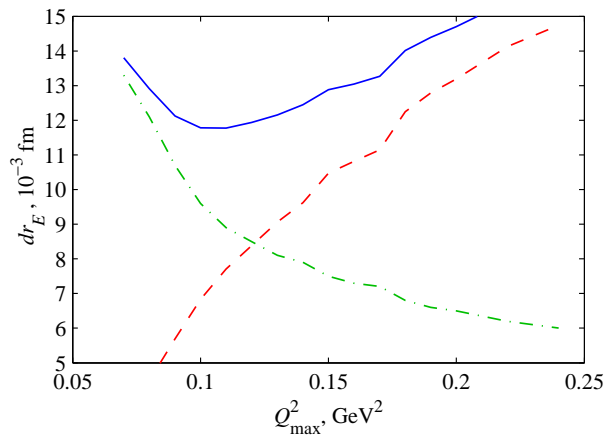}
  \caption{Estimated errors in electric radius vs. $Q_{\rm max}$ for $n=2$,
  statistic (dash-dotted),
  systematic (dashed) and total (solid).}\label{Fig1}
\end{figure}

Thus we suggest that for each $n$ there is some optimal $Q_{\rm max}^2(n)$
with which the total error is minimal. Indeed, this was confirmed
by fitting of pseudo-data. As an example, the dependence of the error on
$Q_{\rm max}^2$ is shown in Fig.~\ref{Fig1} for $n=2$. 
The curves are not smooth because of discrete nature of the data.
The optimal $Q_{\rm max}^2(n)$ values are summarized in Table~\ref{Table1}.
These optimal values are different for electric and magnetic radii.
This means that one does not obtain the best estimates for both $r_E$
and $r_M$ from a single fit; it is necessary to perform two fits
with different $Q_{\rm max}$ --- one for $r_E$ (upper half of the table)
and another for $r_M$ (lower half).
\begin{table}[h]
%
\[ \begin{array}{|c*{2}{|c|c|c|}*2{|c|c|}}
\hline
 \multicolumn{4}{|c||}{$Fit type$} &
 \multicolumn{3}{|c||}{$Errors$} &
 \multicolumn{2}{|c||}{$Results$} &
 \multicolumn{2}{|c|}{$w/o TPE$} \\
\hline
 & n & Q_{\rm max}^2 & $points$ & dr_{\rm stat} & dr_{\rm syst} & dr & r & \chi^2/{\rm d.o.f.} & r & \chi^2/{\rm d.o.f.} \\
\hline
    & 2 & 0.10 & 159 & 0.0096 & 0.0068 & 0.0117 & 0.895 & 0.52 & 0.885 & 0.56 \\
r_E & 3 & 0.30 & 261 & 0.0088 & 0.0066 & 0.0110 & 0.912 & 0.70 & 0.904 & 0.74 \\
    & 4 & 0.98 & 358 & 0.0097 & 0.0054 & 0.0111 & 0.914 & 0.75 & 0.905 & 0.79 \\
\hline
\hline
    & 2 & 0.07 & 126 & 0.0175 & 0.0248 & 0.0304 & 0.836 & 0.50 & 0.803 & 0.52 \\
r_M & 3 & 0.19 & 226 & 0.0129 & 0.0158 & 0.0204 & 0.885 & 0.68 & 0.859 & 0.73 \\
    & 4 & 0.98 & 358 & 0.0100 & 0.0158 & 0.0186 & 0.876 & 0.75 & 0.857 & 0.79 \\
\hline
\end{array} \]
\caption{Fit parameters and results. $Q_{\rm max}^2$ values are in $\GeV^2$,\ \ $r$ and $dr$ --- in fm.}\label{Table1}
\end{table}
\section{Results}
Finally, the fitting procedure was applied to the actual experimental data,
with $n=2..4$ and with optimal $Q_{\rm max}^2$ for each $n$.
The data were taken from
Refs.~\cite{data1,data2,data3,data4,data5,data6,data7,data8,data9}.
The resulting radii and $\chi^2$ per degree of freedom are shown in 
the ``Results'' section of Table~\ref{Table1}.
Choosing different $n$ we obtain somewhat different radii, but they agree
with each other within their estimated uncertainties.
The most precise values are obtained: 
for the electric radius with $n=3$,\quad
$r_E = 0.912 \pm 0.011$ fm, 
for the magnetic radius with $n=4$,\quad
$r_M = 0.876 \pm 0.019$ fm.
The magnetic radius is smaller than the electric one,
but has larger uncertainty
(because the magnetic FF weakly affects low-$Q^2$ cross-section).
In Table~\ref{Table2} our results are compared
to other recent determinations of proton radii.
\begin{table}
\begin{tabular}{|c|c|l|l|}
\hline
  $r_E$      &  $r_M$       & \hfil Ref. \hfil                 & \hfil Method used \hfil\\
\hline
   0.847     &   0.836      & P. Mergell {\it et al.} \cite{MMD} & $eN\to eN$, dispersion analysis\\
0.822..0.835 & 0.843..0.852 & M.A. Belushkin {\it et al.} \cite{MMD2} & $eN\to eN$, dispersion analysis, pQCD approach \\
0.840..0.852 & 0.849..0.859 & M.A. Belushkin {\it et al.} \cite{MMD2} & $eN\to eN$, dispersion analysis, SC approach \\
    ---      &  0.778(29)   & A.V. Volotka {\it et al.} \cite{HFS} & HFS in hydrogen \\
 0.83(5)     &    ---       & I.~Eschrich {\it et al.} \cite{SELEX} & $pe \to ep$ \\
 0.880(15)   &    ---       & R. Rosenfelder \cite{Rosen}  & $ep\to ep$, Coulomb corr. \\
 0.895(18)   &    ---       & I. Sick \cite{Sick}  & $ep\to ep$, Coulomb corr., continued fraction param. \\

 0.912(11)   &  0.876(19)   & This paper & \\
\hline
\end{tabular}
\caption{Recent determinations of proton radii}\label{Table2}
\end{table}

Our best value for electric radius is larger than all others.
However, it is still in agreement with the one obtained in Ref.~\cite{Sick}
using continued fraction parameterization for FFs.
Though our value is somewhat larger,
their difference lies within estimated errors.
Thus the difference does not imply that either value is incorrect,
but gives an impression of the precision to which we know $r_E$.
Nevertheless, there is one interesting implication.
The theoretical value of classic $2s-2p$ Lamb shift with proton size correction
can be written as \cite{LowE}:
\be
 \Delta\nu = 1\,057\,687(2) + 196 r_E^2,
\ee
where $\Delta\nu$ is given in kHz and $r_E$ --- in fm.
With our $r_E$ value this equation yields $\Delta\nu = 1\,057\,850(2)(4)$~kHz,
in agreement with the most precise Lamb shift measurement result,
$1\,057\,857.6(2.1)$~kHz \cite{LowE,Lamb}.

To study the influence of TPE on the obtained results,
we repeat the fit without TPE corrections to data
[putting $\sigma^{2\gamma}=0$ in Eq.(\ref{sigma})].
The results are shown in the last two columns of Table~\ref{Table1}.
We see that $\chi^2$ is slightly bigger in this case, that is,
TPE-uncorrected data are fitted worse.
Both electric and magnetic radius increase after taking TPE into account,
$r_E$ by $\sim 0.01$ fm and $r_M$ by $\sim 0.03$ fm.
The change is comparable with the error for $r_E$,
but is more significant for $r_M$.
This result refutes our claim made in Ref.~\cite{ourLow}, where
we suggest, basing on $Q^2$-dependence of TPE amplitudes,
that the inclusion of TPE should lessen obtained radii.
This is because the data points from typical experiment
do not lie at fixed $\ve$, but lie at fixed energy.
The interplay of $\ve$- and $Q^2$-dependence leads to the increase
of the radii.

Though our work was aimed at the extraction of $r_E$ and $r_M$,
as a by-product we also obtain FF parameterizations in the low-$Q^2$ region.
It may be interesting to compare them to other existing parameterizations,
in particular, to the latest ``global fit'' \cite{paramA}.
\begin{figure}[b]
\parbox{0.47\textwidth}
{ \includegraphics[width=0.45\textwidth]{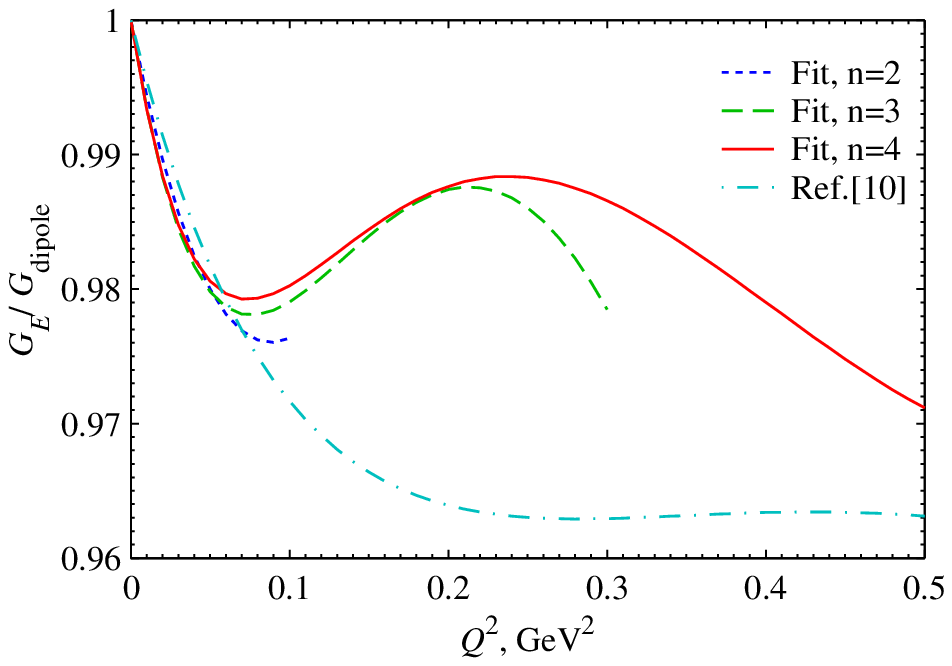}\hfill
  \caption{Electric FF parameterizations, normalized
    by dipole FF $G_{\rm dipole} = (1+Q^2/0.71)^{-2}.$
  }\label{fig:Ge}
}\hfill
\parbox{0.47\textwidth}
{ \includegraphics[width=0.45\textwidth]{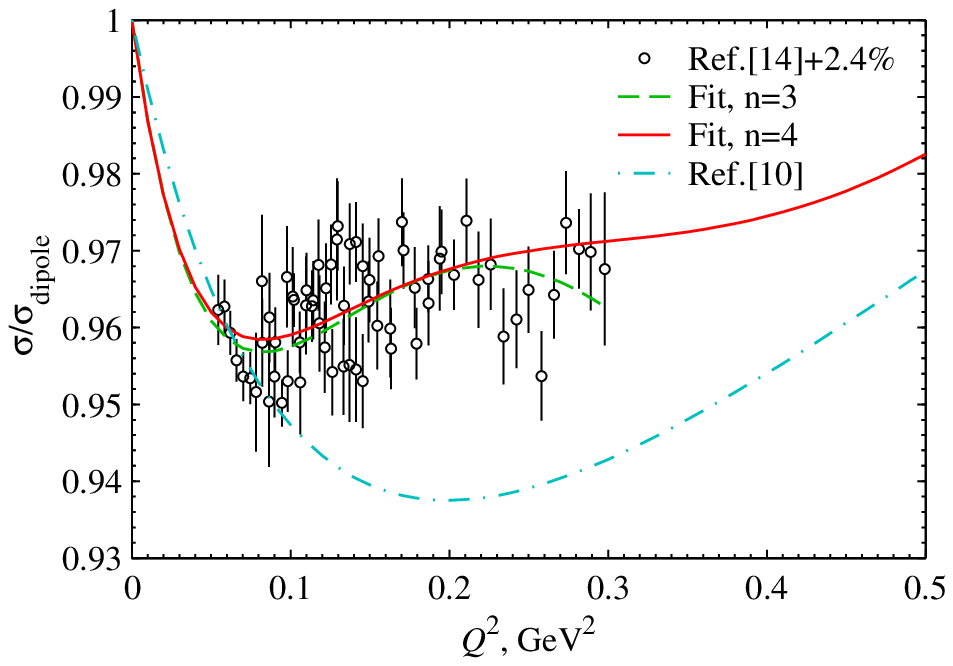}
  \caption{The cross-section at $\ve=1$.
    $\sigma_{\rm dipole}$ is the cross-section
    calculated assuming dipole FFs.
  }\label{fig:bott}
}
\end{figure}
In Fig.~\ref{fig:Ge} we plot the global $G_E$ parameterization
from Ref.~\cite{paramA} and three parameterizations
arising from three ``best fits'' in Table~\ref{Table1}.
Each of them is valid for $Q^2$ less than corresponding $Q_{\rm max}^2$.
All three curves behave similarly and differently
from the fit of Ref.~\cite{paramA}.
After the examination of data sets used in Ref.~\cite{paramA},
we suppose that the difference is due to the omission of data
from Ref.~\cite{data4}, included in the present analysis.
The kinematics of this experiment is:
$\ve \approx 1$ and $Q^2 = 0.05$ to $0.3 \GeV^2$,
thus the data impose important constraints on low-$Q^2$ $G_E$ behaviour.
The parameterization from Ref.~\cite{paramA} totally fails to describe
these data, while in our analysis they were fitted reasonably well
(see Fig.~\ref{fig:bott}).
\begin{figure}
 \includegraphics[width=0.45\textwidth]{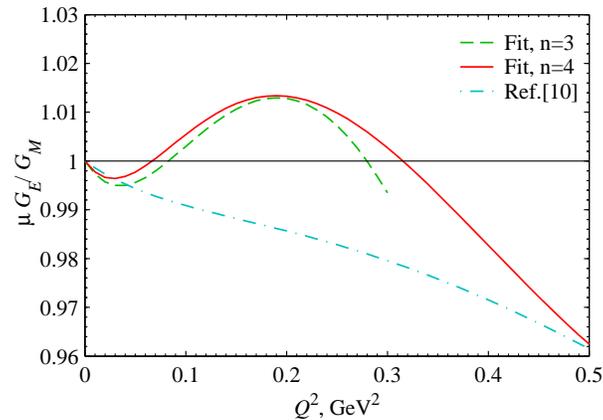}
 \caption{$\mu G_E/G_M$ ratio.}\label{fig:ge/gm}
\end{figure}
In Fig.~\ref{fig:ge/gm} we plot $\mu G_E/G_M$ ratio, derived from our
FF parameterizations. This ratio is unity at $Q^2=0$ and
is known to fall significantly at $Q^2 > 1 \GeV^2$.
However its precise behaviour at low $Q^2$ it not well established.
Our results suggest that this ratio crosses unity at $Q^2 \approx 0.3 \GeV^2$
and slightly grows at lower $Q^2$, returning back to 1 at
$Q^2 \lesssim 0.1 \GeV^2$.
It is interesting to note that similar behaviour may be favoured
by recent polarization transfer measurements \cite{Hig}.
\section*{Acknowledgements}
The author is grateful to A.P.~Kobushkin and J.~Arrington for
valuable comments and discussions.
This work was supported by the Program of Fundamental Researches
of the Department of Physics and Astronomy of National Academy of Sciences, Ukraine.
%

\end{document}